\documentclass{svmult}

\usepackage{makeidx}     
\usepackage{graphicx}    
\usepackage{multicol}    
\usepackage{amsmath,amssymb,pifont}
\usepackage{epsf,rotating,epsfig}
\usepackage{color}
\usepackage{graphicx}

\makeindex             

\definecolor{light}{gray}{0.95}
\definecolor{heavy}{gray}{0.35}
    \def\ie{i.\,e.\ }
    \def\eg{e.\,g.\ }
    \def\pabl#1#2{\frac{{\rm\partial} #1}{{\rm\partial} #2}}
    \def\vgas{\boldsymbol{v}_{\rm gas}}
    \def\Vl{{V_{\ell}}}
    \def\vdrn{{\boldsymbol v}_{\rm dr}^{\hspace{-0.9ex}^{\circ}}}
    
    \def\xx{{\boldsymbol x}}
    \def\Vl{{V_{\ell}}}
    
    \def\nH{n_{\langle{\rm H}\rangle}}
    \def\Nl{{N_{\ell}}}
    \def\platz#1{\rule[-1.8mm]{0mm}{5.5mm}#1}
    \def\er{\boldsymbol{e}_r}

    \def\abl#1#2{\frac{d #1}{d #2}}
    \def\tmix{\tau_{\rm mix}} 
    \def\aquer{\langle a\rangle}

\begin{document}

\title*{The multi-scale dust formation in substellar atmospheres}
\author{Christiane Helling\inst{1,2,3}\and
Rupert Klein\inst{2,4,5}\and Erwin Sedlmayr\inst{1}}
\institute{Zentrum f\"ur Astronomie und Astrophysik, TU Berlin, Hardenbergstra{\ss}e~36, D-10623 
     Berlin\\ \texttt{sedlmayr@astro.physik.tu-berlin.de}
\and Konrad-Zuse-Zentrum f\"ur Informationstechnik Berlin,
  Takustra{\ss}e~7, D-14195 Berlin\\ \texttt{rupert.klein@zib.de}
\and Sterrewacht Leiden, P.O Box 9513, 2300 RA Leiden\\ 
\texttt{helling@strw.leidenuniv.nl}
\and Fachbereich Mathematik und
  Informatik, Freie Universit\"at Berlin, Takustra{\ss}e~7, D-14195
  Berlin 
\and Potsdam Institut f\"ur Klimafolgenforschung,
  Telegrafenberg A31, D-14473 Potsdam}
%
%
\maketitle

\begin{abstract}
  Substellar atmospheres are observed to be irregularly variable for
  which the formation of dust clouds is the most promising candidate
  explanation. The atmospheric gas is convectively unstable and, last
  but not least, colliding convective cells are seen as cause for a
  turbulent fluid field. Since dust formation depends on the local
  properties of the fluid, turbulence influences the dust formation
  process and may even allow the dust formation in an initially
  dust-hostile gas. 
  
  A regime-wise investigation of dust forming substellar atmospheric
  situations reveals that the largest scales are determined by the
  interplay between gravitational settling and convective
  replenishment which results in a dust-stratified atmosphere. The
  regime of small scales is determined by the interaction of turbulent
  fluctuations. Resulting lane-like and curled dust distributions
  combine to larger and larger structures. We compile necessary
  criteria for a subgrid model in the frame of large scale simulations
  as result of our study on small scale turbulence in dust forming
  gases.

\end{abstract}

\section{Introduction}

The astrophysical field of substellar objects -- of which brown dwarfs
shall be considered as an example -- has gained considerable attention
during the last few years since more and more brown dwarfs and
extrasolar planets could be detected. The need for appropriate models
has increased with increasing observational resolution power which,
however, can only provide information about the largest scale
structures. The major challenge has been the observational evidence
for the presence of dust (i.e. small solid or fluid particles) in
substellar atmospheres where energy is mainly transported by
convection. These gas parcels collide after their ascend through the
atmosphere during which they adiabatically expand.  Such adiabatic
collisions cause turbulence, and the turbulent kinetic energy is
transfered cascade-like from large to small scales.  Due to its
chemical nature, the dust formation process itself (like combustion)
depends on local quantities like {\it temperature}, {\it density} and
{\it chemical composition of the gas}, which can vary on much smaller
spatial scale lengths than those accessible by observations. Photon
interactions, molecular collisions and friction influence the dust
formation process and proceed on scales comparable to or even less
than one mean free path of the fluid, which belongs to the smallest
relevant scale regime. The dust formation is hence initiated and
controlled by small-scale processes, but produces consequences on the
largest, observable scales.

The immediate couplings between chemistry, hydrodynamics and
thermodynamics due to the presence of dust can cause an amplification
of initially small perturbations resulting in the formation of
large-scale dust clouds.  The formation of such clouds is one of the
major candidates to explain the observed non-periodic variability
\cite{bm99,bm2001a,bm2001a,bm2001b,mzl2001,gm2002,clark2002} which is
- from the physical point of view - very alike to what we know from
weather-like variation in the Earth atmosphere. However, substellar
object like brown dwarfs and known extraterrestrial planets show more
extreme physical conditions (warmer and hotter or cooler and thinner)
and the direct transfer of Earth's knowledge has to be considered
with care.

It was therefore the aim of this work to investigate the multi-scale
problem of astrophysical dust formation in brown dwarf stars which
requires an appropriate description of the chemical and turbulent
processes and their interactions. The challenge in modeling turbulence
in reactive gas flows lies in an adequate description of {\it all}
relevant scale regimes. This work follows the scale hierarchy from the
smallest to the largest scales. At first, turbulent dust formation
will be studied on small scales (Sect.~\ref{sec:smallscale}) which can
be computationally resolved.  These investigations intend to provide
the basis for a sub-grid closure model for the next (larger) scale
regime where it can be applied and so on. This procedure requires a
detailed, scale-dependent view on the problem \cite{hlsk2001}, or in
other words, needs to adopt different windows of perception
\cite{se2002}.

\section{The model of a dust forming gas flow}\label{sec:model}

Redefining the variables of the system of model equations like $\alpha
\rightarrow \alpha/\alpha_{\rm ref}$ transforms the equations in their
scaled analogues wherein all quantities are dimensionless and can be
compared by number. The non-dimensional characteristic numbers
describe the importance of the source terms and general features can
already be derived by investigating the dimensionless numbers for
physical meaningful reference values.

\medskip\noindent {\underline{\it Complex A:}} The specification of
the source terms in the equations of momentum and energy conservation
results in the following dimensionless system of equations which
describe the fluid field in a substellar atmosphere which is
influenced by the stellar radiation field \cite{holks2001b}:

\begin{eqnarray}
\label{eq:contBD}
(\rho)_t + \nabla \cdot (\rho {\boldsymbol v})&=& 0\\[0.6ex]
\label{eq:motBD}
Sr\,(\rho {\boldsymbol v})_t + \nabla\cdot (\rho {\boldsymbol v}\circ {\boldsymbol
  v}) &=& -\frac{1}{\gamma M^2}\nabla P - \frac{1}{Fr^2}\rho{\boldsymbol g} \\[1.1ex]
\label{eq:enBD}
(\rho e)_t + \nabla \cdot ({\boldsymbol v}[\rho e + P]) &=& Rd_1\,\kappa(T^4_{\rm RE} - T^4),
\end{eqnarray}                                                    
with the caloric equation of state
\begin{equation}
\label{eq:caloBD}
\rho\,e = \gamma M^2\left(\frac{\rho {\boldsymbol v}^2}{2} + \frac{1}{Fr^2}\rho gy\right) + \frac{P}{\gamma - 1}
\end{equation}
with ${\boldsymbol g}=\{0,g,0\}$.  Radiative heating/cooling is
treated by an relaxation ansatz (r.h.s. Eq.~\ref{eq:enBD}).  For the
characteristic numbers $M$, $Fr$, $Sr$, and $Rd_1$, which relate
certain physical processes, see Table~\ref{tab:kennz1}.

\bigskip\noindent \underline{\it Complex B:} \,\, The dust formation
process is considered as a two step process.  At first, seed particles
form out of the gas phase ({\it nucleation}) which provide the first
surfaces. Subsequent growth by surface reactions results in the
formation of (chemically) macroscopic $\mu$m-sized particles.
Nucleation, growth, evaporation, drift and element
depletion/enrichment are physical and chemical processes which occur
simultaneously in an atmospheric gas flow and may be strongly coupled.
Following the classical work of \cite{gs88}, partial differential
equations which describe the evolution of the dust component by means
of moments of its size distribution function, $f(V)$, were derived in
\cite{wh2003a} explicitly allowing for ${\boldsymbol v}_{\rm
  gas}\ne{\boldsymbol v}_{\rm dust}$.  The conservation form of the
dust model equations allows for a fast numerical solution in the frame
of extensive model simulations:

\begin{equation}
  \hspace*{-3mm}
  \pabl{}{t}\big(\rho L_j\big) + \nabla\big(\vgas\,\rho L_j\big) \;= \;
  \underbrace{\int\limits_\Vl^\infty
     \sum\limits_k R_k\,V^{j/3}dV}_{\displaystyle{\cal A}_j}
  \;-\;\underbrace{\nabla\!\int\limits_\Vl^\infty\!
     f(V)\,V^{j/3}\,\vdrn(V)\,dV}_{\displaystyle{\cal B}_j},
  \hspace*{-3mm}
  \label{eq:moment0}
\end{equation}
with $\vdrn(V)$ the grain size dependen equilibrium drift velocity
(for more details please consult \cite{wh2003a}). The $j$-th moment
of the dust size distribution function $L_ j\,{\rm [cm}^j\!/{\rm g]}$
is defined by
\begin{equation}
  \rho L_j(\xx,t) 
  = \int\limits_{\rm \Vl}^{\infty} f(V,\xx,t)\,V^{\rm j/3}\,dV \ .
  \label{eq:MomentDef}
\end{equation}
The source term ${\cal A}_j$ expresses the effects of nucleation and
surface chemical reactions on the dust moments. Compared to the
classical moment equations \cite{gs88}, ${\cal B}_j$ is an additional, advective
term in the new dust moment equations which comprises the effects
caused by a size-dependent drift motion of the grains. The meaning of
the dust variables is summarizes in Table~\ref{tab:DCquantities}.

\begin{table}[h]
\caption[Definition of dust quantities]{Definitions, meanings, and units of the dust and chemical quantities\\ (CE = chemical equilibrium).}
\label{tab:DCquantities}
\vspace*{0.2cm}
\begin{tabular}{cp{1.5cm}|p{6.8cm}}
\hline
Quantity & Unit & Meaning\\
\hline
\multicolumn{3}{c}{\bf Variables}\\
\hline
$\vgas$ & [cm\,s$^{-1}$] &  hydrodynamic gas velocity\\
$\boldsymbol{v}_{\rm dust}$ & [cm\,s$^{-1}$] & hydrodynamic dust velocity\\
$f(V)$ & [cm$^{-6}$] & grain size distribution function\\
$V$  & [cm$^3$] &  volume of the dust particle\\
$R_{\rm k}$ & [s$^{-1}$]   & surface chemical reaction rates\\
$L_{\rm j}$ & [cm$^{\rm j}$\,g$^{-1}$]  & dust moments $(j\, \in\, \mathbb{N})$\\
$\rho L_0 = n_{\rm d}$ & [cm$^{-3}$] & number of dust particles\\
$\sqrt[3]{\frac{3}{4\pi}} \frac{L_1}{L_0} = \langle a\rangle$ & [cm] &   mean radius of dust particles\\
$\sqrt[3]{36\pi} \frac{L_2}{L_0} = \langle A\rangle$ & [cm$^2$] &  mean dust surface\\
$\frac{L_3}{L_0} = \langle V\rangle$ & [cm$^3$] &   mean dust volume\\
$\epsilon_{\rm x}$ & [--] & element abundance relative  to hydrogen\\
$n_{\rm r}$ & [cm$^{-3}$] & number density of gas species $r$\\
$\nH=\frac{\rho}{\!1.427\,{\rm amu}}$   & [cm$^{-3}$] & total hydrogen density\\
$v_{\rm rel, x}$ & [cm\,s$^{-1}$] &  thermal relative velocity\\
                 & &   of the gas species ${\rm x}$\\
\hline
\multicolumn{3}{c}{\bf Constants}\\
\hline
$\nu_{i,0}$ && stoichiometric ratios  of homogeneous nucleation\\
$\nu_{i,r}$ && stoichiometric ratios  of surface reaction $r$\\
\hline
\end{tabular}
\end{table}

\smallskip
\noindent
The motion of a dust grain with a velocity ${\boldsymbol v}_{\rm
dust}$ is determined by an equilibrium between the force of gravity,
${\boldsymbol F}_{\rm grav}$, and the frictional force, ${\boldsymbol
F}_{\rm fric}$, ({\it equilibrium drift} $\,\Rightarrow \vdrn(V)$; see
discussion in \cite{wh2003a}).  Depending on the particle size and the
density of the surrounding fluid, the character of the hydrodynamic
situation changes which also influences the growth process of the
particles. The dimensionless dust moment equations for nucleation,
growth, evaporation, and equilibrium drift write, \eg for the case of
a {\sl subsonic free molecular flow}, are\vspace*{-5mm}

\begin{eqnarray}
  \label{eq:dustlKn}
  \big[{\rm Sr}\big]\big(\rho L_j\big)_t +  
          \nabla ({\vec v}_{\rm gas}\,\rho L_j)&=& \big[{\rm Sr}\cdot{\rm Da^{nuc}_{d}\!\cdot Se_j}\big]\,J(\Vl) 
  \hfill \\[0.2ex] \hspace*{8mm} 
\nonumber
  &+&\,\left[{\rm Sr}\cdot{\rm Da^{gr}_{d, lKn}}\right]\frac{j}{3}\,
     \chi^{\rm net}_{\rm lKn}\,\rho L_{j-1}  \\[0.2ex] \hspace*{8mm} 
\nonumber
  &+&\,\left[\left(\frac{\pi\gamma}{32}\right)^{\!1/2}
     \frac{\rm Sr\cdot M\cdot Dr}{\rm Kn^{^{HD}}\!\cdot Fr}\right]
     \xi_{\rm lKn} \nabla\!\left(\frac{L_{j+1}}{c_T}\,\er\!\right),
   \hspace*{-15mm}\hfill
\end{eqnarray}
\vspace*{-1mm}

\noindent
with $j=0, 1, 2\ldots$ the order of the dust moments.  The first
source term on the r.h.s. describes the seed formation (nucleation),
the second the growth process, and the third describes the transport
of already existing dust particles by drift.  In those cases, where
dust and gas are positionally coupled and drift effects are negligible,
Eq.~(\ref{eq:dustlKn}) simplifies to

\begin{equation}
  \big[{\rm Sr}\big]\big(\rho L_j\big)_t +  
          \nabla ({\vec v}_{\rm gas}\,\rho L_j)= \big[{\rm Sr}\!\cdot {\rm Da^{nuc}_{d}\!\cdot Se_j}\big]\,J(\Vl) 
  + \left[{\rm Sr}\!\cdot {\rm Da^{gr}_{d}}\right]\frac{j}{3}\,
     \chi^{\rm net}\,\rho L_{j-1}.
  \label{eq:dustoD}
\end{equation}
\vspace*{-1mm}

\noindent
For the definition of the characteristic numbers in square brackets
see Table~\ref{tab:kennz1}.

The element depletion is taken into account by evaluating the
consumption of each involved element $i$ with relative abundance to
hydrogen $\epsilon_i$ by nucleation and growth
\begin{multline}
\big[{\rm Sr}\big](\rho \epsilon_{\rm x})_t + \nabla \cdot ({\boldsymbol  v}\,\rho \epsilon_{\rm x}) =
 - \big[{\rm Sr}\!\cdot{\rm El}\big] \sum_{\rm r=1}^{\rm R}  ( \nu^{\rm nuc}_{\rm x, r}\,{\rm Da^{\rm nuc}_{\rm d}} \; N_l J_*\hfill\\ 
+ \nu^{\rm gr}_{\rm x, r}\,{\rm Da^{\rm gr}_{\rm d}}\;\alpha_{\rm r} n_{\rm x, r} v_{\rm rel, x} \, \rho L_2).
\label{eq:BDverbrauch}
\end{multline}
$\epsilon_{\rm x}$ is the element abundance of the chemical element x
(e.g. Ti, Si, O) in mass fractions, which is consumed by nucleation
(first term, r.h.s.) and growth (second term, r.h.s.) with the
stoichiometric factor $\nu_{\rm x, r}$ due to the reaction r.  One
equation needs to be solved for each element involved in the dust
formation process. 

\begin{sidewaystable}
\begin{center}
\caption[Characteristic numbers]{Characteristic numbers and reference values of the scaled model equations in Sect.~\ref{sec:model}.}
\smallskip
\label{tab:kennz1}
\vspace*{0.6mm}
\begin{small} \noindent\hspace*{-0.2cm}
\begin{tabular}{|c|c||c|lr|}
\hline
\multicolumn{2}{|c||}{\bf Characteristic numbers} &
\multicolumn{3}{|c|}{\bf Independent reference values}\\
\multicolumn{2}{|c||}{\bf } &
\multicolumn{3}{|c|}{\bf (fundamental dimensions) }\\
                     
\hline
\platz{Mach number} & ${\rm M }=\frac{v_{\rm ref}}{c_{\rm s}}$ &
density & $\rho_{\rm ref}$  & [g/cm$^{3}$] \\ 

Froude number & ${\rm Fr}=\frac{v^2_{\rm ref}}{l_{\rm ref}}
          \frac{1}{g_{\rm ref}}$ & temperature & $T_{\rm ref}$ & [K]\\

Strouhal number & ${\rm Sr}= \frac{l_{\rm ref}}{t_{\rm ref}v_{\rm ref}}$&
velocity          & $v_{\rm ref}$ & [cm/s] \\

hydrodyn. Knudsen number & ${\bf\rm Kn}^{\rm^{HD}}= 
          \frac{l_{\rm ref}}{2\langle a \rangle_{\rm ref}}$&

length            & $l_{\rm ref}$ & [cm] \\
Knudsen number & ${\rm Kn} = \frac{\bar{\ell}_{\rm ref}}
                      {2\langle a \rangle_{\rm ref}}$&
gravitational acceleration  
                  & $g_{\rm ref}$ & [cm/s$^2$]\\

Drift number & ${\bf\rm Dr}= \frac{\rho_{\rm d, ref}}{\rho_{\rm ref}}$&
nucleation rate   & $J_{\rm ref}/n_{\rm <H>, ref}$ & [s$^{-1}$]\\

1. Radiation number & ${\rm Rd_1} = 4\kappa_{\rm ref}\sigma T^4_{\rm ref} \cdot 
            \frac{t_{\rm ref}}{ P_{\rm ref}}$ & 
 mean particle radius
                  & $\langle a \rangle_{\rm ref}$ & [cm] \\
Sedlma\"yr number $(j \in \mathbb{N}_0)$ & ${\bf\rm Se_{\rm j}} = \left( \frac{a_\ell}
                   {\langle a \rangle_{\rm ref}}\right)^{\rm j}$& 
element abundance& $\epsilon_{\rm ref}$ &\\
\!\!Damk\"ohler no. of nucleation\!\! & ${\bf\rm Da^{\rm nuc}_{\rm d}} =
          \frac {t_{\rm ref} J_{\rm ref} }
          {\rho_{\rm ref}L_{0, {\rm ref}}}$&
&&\\
\!\!\platz{Damk\"ohler no. of growth} (lKn)\!\!&
         ${\bf\rm Da^{\rm gr}_{\rm d, lKn}} =
                    \frac{t_{\rm ref}\chi_{\rm ref, lKn}}
                   {(\frac{4\,\pi}{3}\langle a \rangle^3_{\rm ref})^{1/3}}$ &
&&\\

Element consumption number & ${\rm El}=\frac{\rho_{\rm ref}L_{\rm 0, ref}N_l}{n_{\rm <H>, ref}\epsilon_{\rm ref}}$ &&&\\
\hline
\multicolumn{2}{|c||}{\bf Combined characteristic number} & \multicolumn{3}{c|}{\bf Dependent reference values}\\
\hline
         &&&\\[-2.5ex]
\platz{combined drift number (lKn)}& $\displaystyle\left(\frac{\pi\gamma}{32}\right)^{\!1/2} 
          \frac{\rm Sr\cdot M\cdot Dr}{\rm Kn^{^{HD}}\!\!\cdot Fr}$&
thermal pressure  & $P_{\rm ref} = \frac{\rho_{\rm ref}kT_{\rm ref}} {\bar{\mu}}$
 & [dyn/cm$^2$] \\
&& total hydrogen density &  $n_{\rm <H>, ref}=\rho_{\rm ref}/\sum\epsilon_i m_i$ & [cm$^{-3}$]\\
&& hydrodynamic time    & $t_{\rm ref} = \frac{l_{\rm ref}}{v_{\rm ref}}$ & [s]\\
&& 0$^{\rm th}$~dust moment ($=n_{\rm d}/\rho$)
                  & $L_{0, \rm ref}=\frac{\Delta V_{\rm SiO} 
                    n_{\rm ref, SiO}}{\rho_{\rm ref}\frac{4\pi}{3}
                    \langle a \rangle_{\rm ref}^3}$\,\,($^{*}$)\hspace*{-5mm} & [1/g]\\
&& growth velocity
& $\chi^{\rm ref}_{\rm lKn}$ & [cm/s]\\
&& mean free path
& $\bar{\ell}_{\rm ref}$  & [cm]\\
& & opacity & $\kappa_{\rm ref} = \kappa_{\rm ref}(\sigma^{\rm abs, gas}_{\rm ref}, L_{\rm 3, ref})$ & [cm$^{-1}$]\\
& &  molecular number density & $n_{\rm ref} = n_{\rm ref, SiO}\approx\epsilon_i n_{\rm <H>, ref}$\,\,($^{**}$) & [cm$^{-3}$]\\
\hline
\end{tabular}
\end{small}
\end{center}
\noindent
\hspace*{1cm}
{\scriptsize
\parbox{22cm}{lKn = large Knudsen number ($Kn\gg 1$)\hspace{0.4cm}
($^*$) -- approximation (compare Eq.~\ref{eq:moment0})
\hspace{0.4cm} ($^{**}$) -- to be determined from chemical equilibrium
calculations
\begin{tabbing}
{\bf Parameters:} \=  $\sigma$ - Stefan-Boltzmann constant $\left[\frac{\mbox{erg}}{\mbox{s\,cm$^2$\,K$^4$}}\right]$, $a_l$ - hypothetical monomer radius [cm], $\Delta V_{\rm SiO}$ - monomer volume of SiO$_2$ [cm$^3$],\\
\> $\rho_{\rm d, ref}$ - bulk density [g\,cm$^{-3}$], $m_i$ - element mass [g], $\sigma^{\rm abs}_{\rm ref}$ - gas absorption cross section [cm$^{2}$]

\end{tabbing}
}
}
\end{sidewaystable}

\paragraph*{Characteristic behavior}
Adopting typical values for a brown dwarf atmosphere the following
general characteristics of a dust forming fluid can be drawn (for
definitions see Table~\ref{tab:kennz1}):

\begin{itemize}
\item[--] High {\it Reynolds numbers}, $ Re\approx 10^7\ldots 10^{9}$
  indicate that the viscosity in the brown dwarf atmosphere is too
  small to damp hydrodynamical perturbations and a turbulent
  hydrodynamic field can be expected.

\item[--] Assuming a typical turbulence velocity of the order of one
  tenth of the velocity of sound leads to a {\it Mach number} of
  $M\approx \cal{O}$$(0.1)$ for the whole fluid. Fluctuations on
  smaller scales can have $M\approx \cal{O}$$(10^{-2})$.
  
\item[--] The {\it Froude number} $ Fr =\cal{O}$$(10^{-3}\ldots
  10^{-1})$ shows that the gravity only gains considerable influence
  on the hydrodynamics for scale regimes $l_{\rm ref}\geqslant
  H_{\rho}$. Drift terms are important in the macroscopic regime and
  may be neglected in the small scale regimes.  An analysis of the
  combined characteristic drift number shows that the drift term in
  the dust moment equations is mainly influenced by the gravity and
  the bulk density of the grains.

\item[--] The {\it characteristic number for the radiative heating /
    cooling}, $ Rd_1=4 \kappa_{\rm ref}\sigma T^4_{\rm ref} \cdot
  \frac{t_{\rm ref}}{ P_{\rm ref}}$ gives the ratio of the
  radiative and the thermal energy content of a fluid.  The scaling of
  the system influences $Rd_1$ by the reference time $t_{\rm ref}$ which
  increases with increasing spatial scales.

\item[--] An analysis of the characteristic numbers in front of the
  source terms of the dust moment equations reveals a clear
  hierarchy of the processes of dust formation: nucleation $\to$
  growth $\to$ drift.

\end{itemize}

\begin{center}
  \colorbox{light}{\parbox{0.95\textwidth}{\it The governing equations of
      the model problem are those of an inviscid, compressible fluid
      which are coupled to stiff dust moment equations and an almost singular
      radiative energy relaxation if dust is present. The coupling
      between the processes becomes stronger with increasing time
      scales.}}
\end{center}

\bigskip
\noindent
If additionally $\boldsymbol{v}_{\rm gas}\!\to0$, \ie the dust-forming
system reaches the static case, and $t\rightarrow\infty$, the source
terms in Eqs.  (\ref{eq:dustlKn}) must balance each other. In the
$S\!>\!1$ case ($S$ - supersaturation ratio), this means that the gain
of dust by nucleation and growth must be balanced by the loss of dust
by rain-out. In the $S\!<\!1$ case, just the opposite is true, \ie the
loss by evaporation must be balanced by the gain of dust particles
raining in from above.  Both {\it control mechanisms} (in the static
limit) result in an efficient transport of condensible elements from
the cool upper layers into the warm inner layers, which cannot last
forever.
\begin{center}
  \colorbox{light}{\parbox{0.95\textwidth}{\it If the brown dwarf's
      atmosphere is truly static for a long time, there is no other
      than the trivial solution for Eqs.~(\ref{eq:dustlKn}) where the
      gas is saturated ($S\!\equiv\!1$) and dust-free
      ($L_j\!\equiv\!0$).}}
\end{center}

\section{Dust formation on small scales}
\label{sec:smallscale}

\subsection{Numerical approach}
The fully time-dependent solution of the model equations
(Eqs.~\ref{eq:contBD}
--~\ref{eq:enBD},~\ref{eq:dustoD},~\ref{eq:BDverbrauch}) has been
obtained in the small scale regimes by applying a multi-dimensional
hydro code \cite{smk97} which has been extended in order to treat the
complex of dust formation and elemental conservation.

\paragraph*{Initial conditions:}
The initial conditions have been chosen as homogeneous, static,
adiabatic, and dust free, \ie $\rho_0=1$, $p_0=1$, $u_0=0$,
$L_0=0\,(\Rightarrow\, L_j=0)$ in order to represent a (semi-)static,
dust-hostile part of the substellar atmosphere. This allows us to
study the influence of our variable boundaries on the evolution of the
dust complex without a possible intersection with the initial
conditions.

\paragraph*{Boundary conditions and turbulence driving:} 
The Cartesian grid is divided in the cells of the test volume (inside)
and the ghost cells which surround the test volume (outside). The
state of each ghost cell is prescribed by our adiabatic model of
driven turbulence for each time.  The hydro code solves the model
equation in each cell (test volume + ghost cells) and the prescribed
fluctuations in the ghost cells are transported into the test volume
by the nature of the HD equations.  The numerical boundary occurs
between the ghost cells and the initially homogeneous test volume and
are determined by the solution of the Riemann problem.  Material can
flow into the test volume and can leave the test volume.  The solution
of the model problem is considered inside the test volume.

\subsubsection*{Stiff coupling of dust and radiative heating / cooling}

Dust formation occurs on much shorter time scales than the
hydrodynamic processes. Approaching regimes of larger and larger
scales makes this problem more and more crucial.  Therefore, the dust
moment and element conservation equations ({\it Complex B}) are solved
applying an ODE solver in the framework of the operator splitting
method assuming $T, \rho=$const during ODE solution. In
\cite{holks2001b} we have used the CVODE solver (Cohen \& Hindmarsh
2000; LLNL) which turned out to be insufficient for the mesoscopic
scale regime.  CVODE failed to solve our model equations after the
dust had reached its steady state.
\paragraph*{The LIMEX solver:}\label{sec:LIMEX}
The solution of the equilibrium situation of the dust complex is
essential for our investigation since it describes the stationary case of
{\it Complex B} when no further dust formation takes place. The reason
may be that all available gaseous material has been consumed and the
supersaturation rate $S=1$ or the thermodynamic conditions do not
allow the formation of dust. The first case involves an asymptotic
approach of the gaseous number density of $S=1$ which often is
difficult to be solved by an ODE solver due to the choice of too large
time steps. However, the asymptotic behavior is influenced by the
temperature evolution of the gas/dust mixture which in our model is
influenced by radiative heating/cooling. Since the radiative heating/cooling ($Q_{\rm rad} =
Rd\,\kappa(T^4_{\rm RE} - T^4)$ heating/cooling rate) depends on the
absorption coefficient $\kappa$ of the gas/dust mixture which strongly
changes if dust forms.  Consequently, the radiative heating/cooling
rate is strongly coupled to the dust complex which in turn depends
sensitively on the local temperature which is influenced by the
radiative heating/cooling.  It was therefore necessary to include also
the radiative heating/cooling source term in the separate ODE
treatment for which we adopted the LIMEX DAE solver.

LIMEX \cite{dw87} is a solver for linearly implicit systems of
differential algebraic equations. It is an extrapolation method based
on a linearly implicit Euler discretization and is equipped with a
sophisticated order and stepsize control \cite{deu83}. In contrast to
the widely used multistep methods, \eg OVODE, only linear systems of
equations and no non-linear systems have to be solved internally.
Various methods for linear system solution are incorporated, \eg full
and band mode, general sparse direct mode and iterative solution with
preconditioning. The method has shown to be very efficient and robust
in several fields of challenging applications in numerical
\cite{neo98,enod99} and astrophysical science \cite{stra2002}.

\subsection{Microscopic regime}\label{ss:micro}

The smallest scale involved into the structure formation of a
substellar atmosphere is the monomer volume, $\Delta V$, by which a
chemical reaction $r$ increases the volume of a grain. For radiative
transfer and drift effects, the grain size becomes important. On small
hydrodynamic scales, drift effects are negliggibaly small. However,
the aim is to investigate the role of turbulence in the formation of
dust clouds and the possibly related variability of brown dwarfs.
Therefore, the interesting hydrodynamic phenomena in the microscopic
scale regime are acoustic waves. 

\begin{figure}[hb]
  \begin{center}
  \epsfig{file=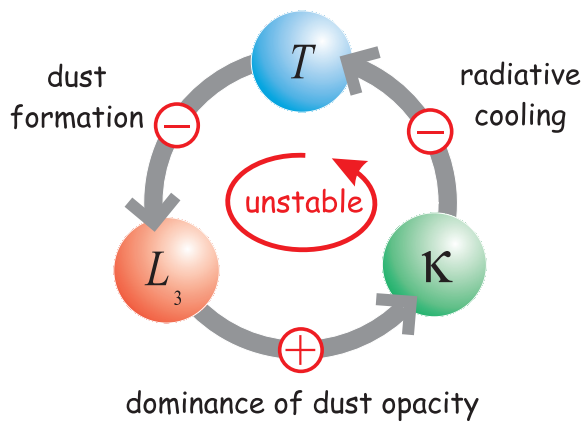, scale=0.8}
  \caption[Feedback loop]{Unstable feedback loop in dust forming
    systems: Dust, once formed, increases the total opacity which
    intensifies radiative cooling. The temperature decreases and
    enters the dust formation window. Therefore, more dust is formed
    which increases the opacity further. An instability establishes
    where the dust improves the conditions for its own formation
    \cite{hel2003}.}
  \label{fig:instabil}
  \end{center}
  \vspace*{-2mm}
\end{figure}

\smallskip\noindent Carrying out numerical simulations of the whole
time-dependent system of model equations (Eqs.~\ref{eq:contBD}
--~\ref{eq:enBD},~\ref{eq:dustoD},~\ref{eq:BDverbrauch}), turbulence
is seen as acoustic waves being by-products of colliding convective
cells \cite{holks2001b} which undergo an adiabatic increase of size
during their upward travel through the substellar atmosphere. A {\it
  feedback loop} establishes in a dust-forming system in which
interacting expansion waves play a key role (Fig.~\ref{fig:instabil}):
The interaction of small-scale disturbances of the fluid field can
cause a local and temporarily limited temperature decrease low enough
to initiate dust nucleation.  These seed particles grow until they
reach a size where the dust opacity is large enough to accelerate
radiative cooling which causes the temperature to decrease again below
a nucleation threshold ($T_{\rm s}$).  Dust nucleation is henceforth
re-initiated which results in a further intensified radiative cooling.
The nucleation rate and consequently also the amount of dust particles
increase further.  This {\it run-away} process is stopped if either
the radiative equilibrium temperature of the gas is reached or all
condensible material is consumed.  Meanwhile, the seed particles have
grown to macroscopic sizes.

The result is a highly variable dust distribution in space and time on
the microscopic scales investigated (\eg $\Delta l \approx 10^2$\,cm,
$\Delta t\approx 0.5$\,s) due to the occurrence of singular nucleation
events. It is immediately clear, that such scales are not easily
resolvable by observations, but these are the scales on which
structure formation is likely to be initiated. However, such small
scale pattern might move and successively initiate dust formation at
various sites in the atmosphere. A larger and larger cloud structure
may thereby form which is, hence, determined by a non-local
hydrodynamic coupling whereas at each site the above outlined local
feedback loop will act (compare Fig.~\ref{fig:vorticity}).

\subsection{Mesoscopic regime}\label{ss:meso}

Following Kolmogoroff's idea, the mesoscopic scales shall be
considered as those where energy is only transfered through the
turbulence cascade, but is not injected or dissipated.  Convective
elements are generated as long as the Schwarzschild criteria is
fulfilled in the substellar atmosphere. The inertia of mass will drive
the convective elements beyond Schwarzschild's boundary, a phenomenon
usually named as overshooting, which extends this zone further. The
mass elements interact, collide, and transfer part of their energy
into small scale structures, thereby producing a whole spectrum of
them being viewed as waves or -- turbulence elements. These waves
propagate and enter even atmospheric regions which are not influenced by
convective motions any more.

The question arises how the dust formation process takes place in such
a stochastically fluctuating thermo- and hydrodynamic environment, and
therefore which influence such small scale disturbances may have on
the whole atmospheric structure and the formation of large scale
pattern possibly responsible for observed variabilities.  A model for
driven turbulence was proposed \cite{hkws2002,hkwns2003} which
simulates a constantly occurring, small-scale energy input assumed to
originate from a convectively active zone. Following Reynolds scale
separation ansatz a background field $\alpha_0({\boldsymbol x}, t)$ is
disturbed by a fluctuation $\delta\alpha({\boldsymbol x}, t)$ of some
variable $\alpha({\boldsymbol x}, t)$ such that

\begin{equation}
\alpha({\boldsymbol x}, t) = \alpha_0({\boldsymbol x}, t)  + \delta\alpha({\boldsymbol x}, t).
\end{equation}
$\alpha(\boldsymbol{x},t)\, \epsilon \, \{
\boldsymbol{u}(\boldsymbol{x},t), P(\boldsymbol{x},t),
S(\boldsymbol{x},t) \}$ ($\boldsymbol{u}(\boldsymbol{x},t)$ -
velocity, $P$ - pressure, $S$ - entropy).  The velocity fluctuation
$\delta \boldsymbol{u}({\boldsymbol x}, t)$ follows the Kolmogoroff
spectrum in $k$-space in which a whole range of wavenumbers is excited
($k_{\rm min}=2\pi/l_{\rm max}$, $k_{\rm max}=2\pi/(3\Delta x)$,
$l_{\rm max}=5\times 10^4$cm -- maximum scale considered, $\Delta x =
10^2$cm -- spatial grid resolution).  The model for driven turbulence
relies on a superposition ansatz of different wave modes and therefore
allows again to carry out direct simulation of the time and space
evolution of the dust complex, now in a stochastically excited medium
typical for the elswise dust-hostile part of a brown dwarf atmosphere.

\begin{figure}[ht]
  \begin{center}
\epsfig{file=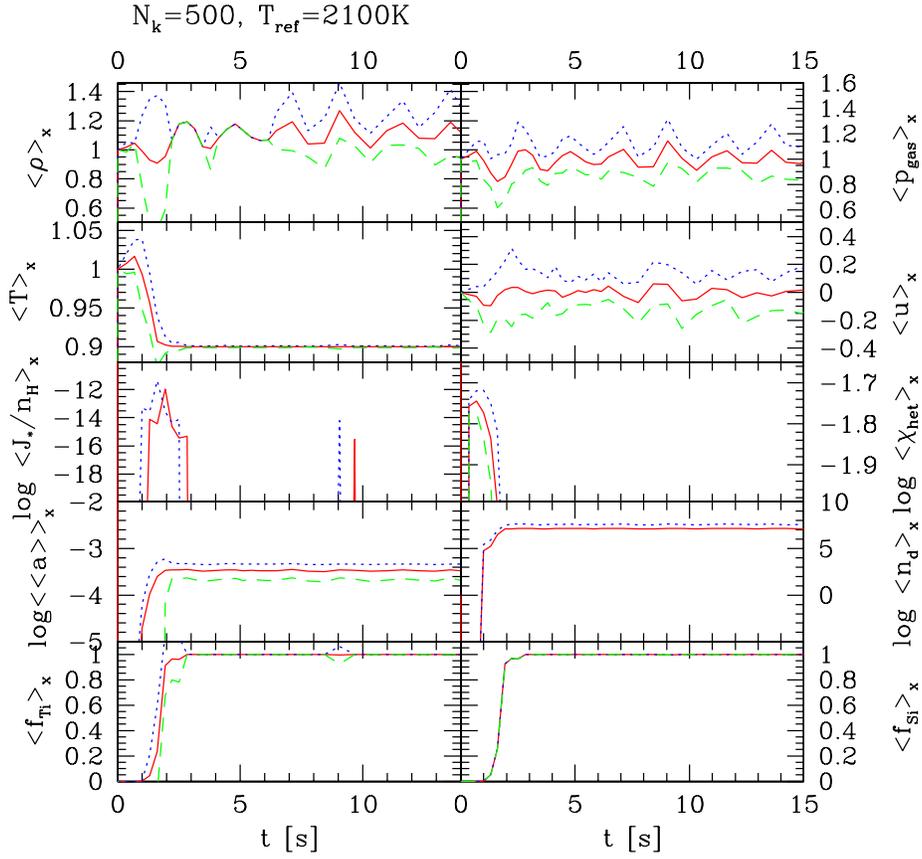, scale=0.7}
\caption[Space-means and standard deviations]{Fluctuating dust
  component in substellar atmospheres. The space-means
  $\langle\alpha(t)\rangle_x$ (solid/red) with the apparent standard
  deviations $\sigma_{\rm N_x-1}^{\alpha}(t)$ (dotted/dashed) as
  function of time for $T_{\rm ref}=2100$K, $\rho_{\rm
    ref}=3.16\,10^{-4}$g\,cm$^{-3}$, $v_{\rm ref}=c_{\rm S}/10$.
  ($\langle\alpha(t)\rangle_x+\sigma_{\rm N_x-1}^{\alpha}(t)$ -
  dotted/blue; $\langle\alpha(t)\rangle_x-\sigma_{\rm
    N_x-1}^{\alpha}(t)$ - dashed/green).}
\label{fig:1Dstoch2100mitStandartdeviation}
  \end{center}
  \vspace*{-2mm}
\end{figure}

\paragraph*{Nucleation events and nucleation fronts}
An initially hydrodynamically homogeneous and dust-free, 500m-sized
gas parcel has been considered in a deep layer of a brown dwarf
atmosphere initially too hot for nucleation. The gas is disturbed by
superimposed waves entering through its boundaries. If
these disturbances carry already a temperature  below the
nucleation threshold ($T_{\rm s}$ - nucleation threshold temperature),
a {\it nucleation front} will develop: The nucleation peak moves
inwards together with the wave and initiates the feedback loop already
known from the microscopic regime by leaving behind first dust seeds.
If the entering temperature disturbances is not enough to cross $T_{\rm
  s}$, interaction with some expansion wave coming from another
direction at some time and some site will take place thereby causing
singular {\it nucleation events} to occur.

A nucleation front tends to homogeneously fill the gas parcel with
dust in 1D situations while nucleation events cause a more
heterogeneous dust distribution due to their short life time.
In more than 1D, also nucleation fronts cause a heterogeneous dust
distribution because their 'parent' waves will interact and influence
the conditions for dust formation.

Large variations in the dust quantities occur when the formation
process starts to influence the hydro- and thermodynamics of the gas
parcel. If all available material has been consumed, the mean dust
quantities reach an almost constant values
(Fig.~\ref{fig:1Dstoch2100mitStandartdeviation}) at a certain place in
the atmosphere which is characterized by $T_{\rm ref}$. During this
active time ($\Delta t \approx 3\,$s), the variation in the number of
dust particles is $\cal{O}$$(10^5\,$cm$^{-3})$ and in the mean
particle size $\cal{O}$$(10^2\,\mu$m).  The variation of the density
is considerable (see standard deviations in
Fig.~\ref{fig:1Dstoch2100mitStandartdeviation}) during the time of
temperature decrease which assures the establishment of the pressure
equilibrium.

\begin{center}
   \colorbox{light}{\parbox{0.95\textwidth}{\it A convectively ascending, initially dust free
    gas element can be excited to form dust by waves running through it. A
    cloud can therefore be fully condensed at much higher temperatures
    than classically expected, \ie in an undisturbed case.}}
\end{center}

\begin{figure}
\vspace*{0.5cm}
\hspace*{-0.2cm}
\epsfig{file=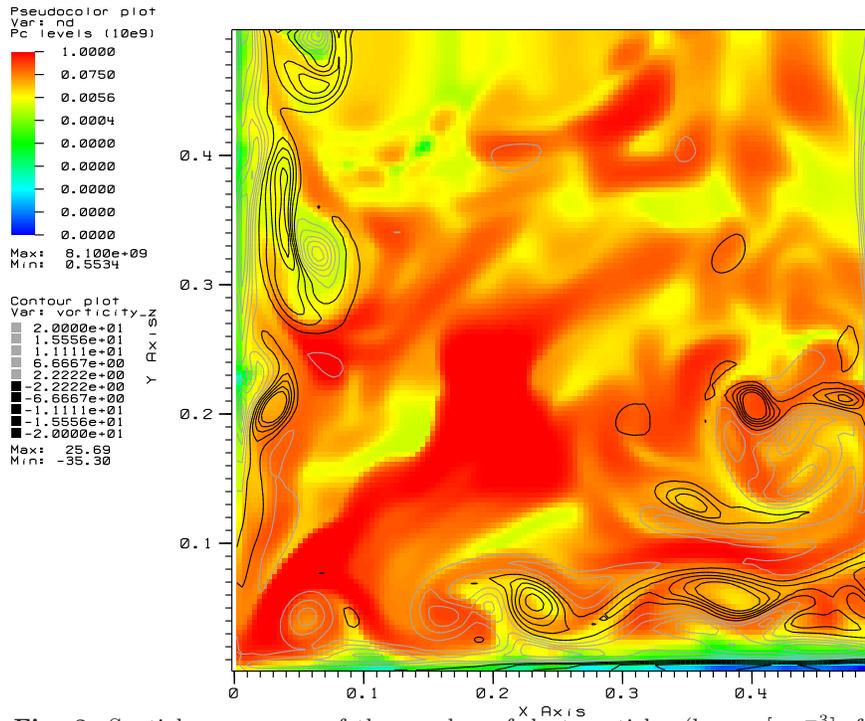, scale=0.6}\\[3.5cm]
\caption[]{Spatial appearance of the number of dust particles ($\log n_{\rm d}$ [cm$^{-3}$]; false color background) and the vorticity ($\nabla\times \boldsymbol{v}$; black and grey contour lines) 
of the 2D velocity field for $t=0.8$s of a simulation with $T_{\rm ref}=2100$\,K, $\rho_{\rm ref}=3.16\,10^{-4}$g\,cm$^{-3}$, $v_{\rm ref}=c_{\rm S}$.}
\label{fig:vorticity}
\end{figure}

\paragraph*{Formation of large scale structures in 2D}
2D simulations with the smallest eddy size being $\lambda^{\rm
  2D}_{\rm min}=5\,$m, the largest are of the size of the gas box
$\lambda^{\rm 2D}_{\rm max}=500\,$m, reveal a very intermittent
distribution of the dust in space during the time of efficient dust
formation (Fig.~\ref{fig:vorticity}).  The very inhomogeneous
appearance of the dust complex is a result of nucleation fronts and
nucleation events comparable to the 1D results.  The nucleation is now
triggered by the interaction of eddies coming from different
directions. Large amounts of dust are formed and appear to be present
in lane-like structures (large $\log\,$n$_{\rm d}$; dark/red areas in
Fig.~\ref{fig:vorticity}). The lanes are shaped by the constantly
inward traveling waves.  Our simulations show that some of the small
scale structure merge thereby supporting the formation of lanes and
later on even larger structures.

Dust is also present in curled structure which indicates the
formation of vortices due to the 2D waves driving turbulence.  As the
time proceeds in the 2D simulation, vortices develop orthogonal to the
velocity field which show a higher vorticity ($\nabla \times
\boldsymbol{u}(\boldsymbol{x}, t)$) than the majority of the
background fluid field. Comparing the distribution of dust particles
(false color background in Fig.~\ref{fig:vorticity}) with the vorticity
(contour lines in Fig.~\ref{fig:vorticity}), shows that the vortices
with high vorticity preferentially occur in dust free regions or
regions with only little amounts of dust present. These vortices can
efficiently drag the dust into regions with still material available
for further condensation shaping thereby larger and larger structures.

\section{Dust formation on large scales}
\label{sec:largescale}

The atmospheric extension ($\approx 100\,$km -- several pressure scale
heights $H_p$) proposes a natural upper limit for the macroscopic
scale regime. On this scale, the interplay between convective
overshooting and gravitational settling (drift) is a major mechanism which
influence the dust structure of the whole atmosphere.  Since the dust
is strongly coupled to the hydro- and thermodynamics of the
atmosphere, gravitational settling and convective overshooting
indirectly alter the atmospheric density and temperature structure
which are important for the spectral appearance of the objects. The
large scale structures are those which are usually accessible by
observations.  Therefore, the understanding and modeling of processes
directly connected with macroscopic scale motions are of particular
interest.

\subsection{The model of a quasi-static dust layer}\label{ssec:modelquasistat}
A theoretical consistent description of dust
formation, gravitational settling, and element depletion
(Eqs.~\ref{eq:dustlKn},~\ref{eq:BDverbrauch}) was needed in order to
understand the formation and the structure of dust cloud layers which
is urgently needed for explaining the spectral appearance of Brown
Dwarfs \cite{wh2003a}. A first application to the quasi-static case of
a brown dwarf atmosphere calculation was carried out by providing a
model of a quasi-static cloud layer, i.e. ${\boldsymbol v}_{\rm
  gas}=0$ and $\frac{\partial L_j}{\partial t}=0$ in
Eqs.~(\ref{eq:contBD})
--~(\ref{eq:enBD}),~(\ref{eq:dustlKn}),~(\ref{eq:BDverbrauch})
\cite{wh2003b}. However, if ${\boldsymbol v}_{\rm gas}=0$
Eqs.~(\ref{eq:dustlKn}) have the trivial solution $L_j\!\equiv\!0$ (see
Sect.~\ref{sec:model}). The physical interpretation of this solution
is that dust grains have once formed in the sufficiently cool layers,
have consumed all available condensible elements up to the saturation
level, and have finally left the model volume by gravitational
settling. Consequently, a truly static atmosphere must be dust-free
which -- in this generality -- contradicts the observations.

This picture changes, however, if we take into account a mixing of the
atmosphere caused by the convection. This mixing leads to an ongoing
replenishment of the gas with fresh, uncondensed matter from the deep
interior of the brown dwarf, which can counterbalance the loss of
condensible elements via the formation and gravitational settling of
dust grains. Thus, a convective mixing can maintain a stationary
situation:\,\ Seed particles nucleate in metal-rich, \ie
supersaturated upwinds. Dust particles grow on top of these nuclei by
the accretion of molecules and rain out as soon as they have reached a
certain size. The sinking grains will finally reach deeper atmospheric
layers which are hot enough to cause their thermal evaporation, which
completes the life cycle of a dust grain in a brown dwarf atmosphere.
We therefore enlarge the static case of our model by a simple
description of the convective mixing such that the stationary dust
moment equations write
\begin{equation}
  -\abl{}{z}\!\!\left(\frac{L_{j+1}}{c_T}\right) \;=\;
  \frac{1}{\xi_{\rm lKn}} \left(
        - \frac{\rho L_j}{\tmix} + \Vl^{\,j/3} J_\star 
        + \frac{j}{3} \chi_{\rm lKn}^{\rm net} \, \rho L_{j-1} \right) 
  \label{eq:dust_static_mixed}
\end{equation}
with algebraic auxiliary conditions
\begin{equation}
  \frac{\nH (\epsilon_i^0-\epsilon_i)}{\tmix} \;=\;
  \nu_{i,0}\,\Nl\,J_\star  
  \,+\, \sqrt[3]{36\pi}\,\rho L_2 \sum\limits_{r=1}^R 
        \nu_{i,r} n_r v^{\rm rel}_r \alpha_r \left(1-\frac{1}{S_r}\right)
        \ .
  \label{eq:elem_static_mixed}
\end{equation}
According to this approach, the gas/dust mixture in the atmosphere is
continuously replaced by dust-free gas of solar abundances on a
depth-dependent mixing time-scale $\tmix(z)$, which can be adapted to
a convection model (see discussion in \cite{wh2003b,wh2003c}).
$\epsilon_i$ is the actual abundance of element $i$ in the gas phase
and $\epsilon_i^0$ its solar value \cite{ag89}. By fitting the mass
exchange frequency of 3D dynamic simulations of surface convection by
\cite{lah2002} with an exponential we can apply a rough description of
$\tmix(z)$.

\subsection{Numerical approach}

Equation~(\ref{eq:dust_static_mixed}) is a system of ordinary differential
equations of first order which can be solved by standard numerical
methods.  The differential equations are integrated inward by means
of the variable transformation $z'\!=\!z_{\rm max}-z$ using the {\sc
  Radau}\,5\,--\,solver for stiff ordinary differential equations
\cite{hw91a}).  For test purposes, the
equation of hydrostatic equilibrium $dP_{\rm gas}/dz = -\rho g$ is
solved in addition to the three moment
equations~(\ref{eq:dust_static_mixed}), where the actual gas pressure
$P\!=\!\sum_i n_i kT$ results from the chemical equilibrium
calculations. The integration is stopped as soon as one of the dust
moments becomes negative, indicating that the dust has completely
evaporated.

\subsection{Structure of a quasi-static cloud layer}\label{ssec:strucquasistat}

\begin{figure}
  \begin{center}
  \epsfig{file=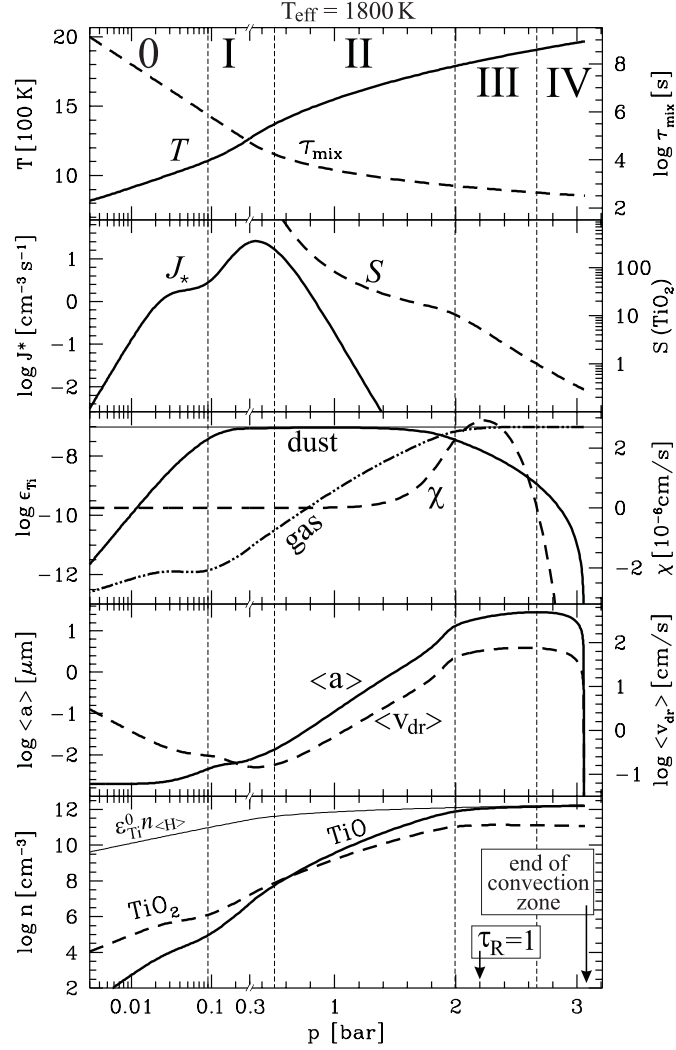, scale=0.69}
  \caption{Calculated TiO$_2$ cloud structures for models with $\log\,g=5$ 
    and $T_{\rm eff}=1800\,$K. Note the scaling of the partly
    logarithmic and partly linear $p$-axis. {\bf 1$^{\bf st}$\,panel:}
    prescribed gas temperature $T$ (solid), according to Tsuji (2002),
    and mixing time-scale $\tmix$ (dashed).  {\bf 2$^{\bf
        nd}$\,panel:} nucleation rate $J_\star$ (solid) and
    supersaturation ratio $S_{\rm TiO_2}$ (dashed). {\bf 3$^{\bf
        rd}$\,panel:} Ti abundance in the dust phase $\epsilon^{\rm
      dust}_{\rm Ti}$ (solid) and in the gas phase $\epsilon_{\rm Ti}$
    (dashed-dotted). The solar value $\epsilon_{\rm Ti}^0$ is
    additionally indicated by a thin straight line. The dashed line
    shows the growth velocity $\chi^{\rm net}_{\rm lKn}$.  {\bf
      4$^{\bf th}$\,panel:} mean particle size
    $\aquer\!=\!\sqrt[3]{3/(4\pi)}\,L_1/L_0$ (solid) and mean drift
    velocity $\langle v_{\rm dr}\rangle\!=\!  \sqrt\pi g \rho_{\rm d}
    \langle a \rangle\,/\,(2\rho c_T)$ (dashed, see Eq.~(66) in
    Paper~II). {\bf 5$^{\bf th}$\,panel:} molecular particle densities
    of TiO (solid) and TiO$_2$ (dashed). For comparison, the
    hypothetical total titanium nuclei density for solar abundances
    $\nH\epsilon^0_{\rm Ti}$ is depicted by a thin solid line.}
  \label{fig:Teff1800}
  \end{center}
  \vspace*{-3mm}
\end{figure}

The vertical structure of the cloud layer results from a competition
between the four relevant physical processes: mixing, nucleation,
growth/evaporation and drift.  Following the cloud structures inward
(from the left to the right in Fig.~\ref{fig:Teff1800}) roughly five
different regions can be distinguished, marked by the Roman digits,
which are characterized by different leading processes concerning the
dust component. 

\smallskip\noindent \underline{\sc {\bf 0.} Dust-poor depleted gas:} High
above the convection zone, the mixing time-scale is large and the
elemental replenishment of the gas is too slow to allow for
considerable amounts of dust to be present in the atmosphere. The few
particles forming here are very small $\aquer\!<\!10^{-2.5}\,\mu$m and
have drift velocities $\langle v_{\rm dr}\rangle \approx 1\,$mm/s to
1\,cm/s which causes these atmospheric layers to become dust-poor.
The gas phase is strongly depleted in condensible elements.  The Ti
abundance in the gas phase $\epsilon_{\rm Ti}$ is reduced by a factor
between $10^5$ and $10^7$ from its solar value.  However, phase
equilibrium ($S\!=\!1$) is not achieved because even the very small
disturbance of the atmosphere by mixing is sufficient to produce a
solution which differs significantly from the trivial solution in the
truly static case.

\smallskip\noindent \underline{\sc {\bf I.} Region of efficient nucleation:}
Nucleation takes place mainly in the upper parts of the cloud layer,
where the temperatures are sufficiently low and the elemental
replenishment by mixing is sufficiently effective. Although the gas is
strongly depleted in heavy elements in these layers, it is
nevertheless highly supersaturated ($S\!>\!1000$) such that
homogeneous nucleation can take place efficiently.  Since very many
seed particles are produced in this way, the dust grains remain small
$\langle a\rangle\!<\!0.01\,\mu$m and have small mean drift velocities
$\langle v_{\rm dr}\rangle \approx 0.1\,$mm/s ... $1\,$mm/s, which are
even smaller than in region~0 because of the higher gas densities.

\smallskip\noindent \underline{\sc {\bf II.} Dust growth region:} With the
inward increasing temperature, the supersaturation ratio $S$ decreases
exponentially which leads to a drastic decrease of the nucleation rate
$J_\star$. Consequently, nucleation becomes unimportant at some point,
\ie {\sl the in-situ formation of dust particles becomes inefficient
  in region~II}.  Here, the condensible elements mixed up by
convective overshoot are mainly consumed by the growth of already
existing particles, which have formed in region~I and have drifted
into region~II. The gas is still strongly supersaturated $S\!\gg\!1$,
indicating that the growth process remains incomplete, \ie the
condensible elements provided by the mixing are not exhaustively
consumed by growth. The dust component in region~II is characterized
by an almost constant degree of condensation ($\propto\!\epsilon^{\rm
  dust}_{\rm Ti}\!\propto\!L_3\!\approx\!{\rm const}$), while the mean
particle size $\aquer$ and the mean drift velocity $\langle v_{\rm
  dr}\rangle$ increase inward. Consequently, the total number of dust
particles per mass ($L_0$) and their total surface per mass
($\propto\!L_2$) decrease.

\smallskip\noindent \underline{\sc {\bf III.} Drift dominated region:} With
the decreasing total surface of the dust particles ($\propto\!L_2$),
the consumption of condensible elements from the gas phase via dust
growth becomes less effective. At the same time, due to the increase
of the mean particle size $\aquer$, the drift velocities increase.
When the dust particles have reached a certain critical size,
$\aquer_{\rm cr}\!\approx\!15\,\mu$m to $50\,\mu$m, the drift becomes
more important than the growth, and the qualitative behavior of the
dust component changes.  This happens at the borderline between
region~II and III, which we denote by rain edge. Although the gas is
still highly supersaturated $S\!\approx\!1\,...\,10$, the in-situ
formation of dust grains is ineffective as in region~II.  The
depletion of the gas phase vanishes in region~III, \ie the gas
abundances approach close-to-solar values. The grains reach their
maximum radii at the lower boundary of region~III (the cloud base):
$\aquer\approx\!30\,\mu$m to $400\,\mu$m at maximum drift velocities.

\smallskip\noindent \underline{\sc {\bf IV.} Evaporating grains:} The
gravitationally settling dust particles finally cross the cloud base
and sink into the undersaturated gas situated below, where $S\!<\!1$.
Here, the dust grains raining in from above evaporate thermally. The
evaporation of these dust particles, however, does not take place
instantaneously, but produces a spatially extended evaporation
region~IV with a thickness of about 1\,km. With decreasing altitude,
the particles get smaller $d\langle a\rangle/dz\!<\!0$ and slower
$d\langle v_{\rm dr}\rangle/dz\!<\!0$.  Consequently, their residence
times increase, and a run-away process sets in which finally produces
a very steep decrease of the degree of condensation, terminated by the
point where even the biggest particles have evaporated completely. We
note that region~IV, in particular, cannot be understood by stability
arguments, but requires a kinetic treatment of the dust complex.

\section{Conclusions}
The dust formation in brown dwarfs atmospheres is a multi-scale
problem where different physical mechanisms characterize the different
scale regimes:
\begin{itemize}
\item[a)] In the {\it microscopic scale regime}, single acoustic waves
  interact, thereby initiate dust formation.
\item[b)] The superposition of a whole spectrum of
  turbulence elements determine the {\it mesoscopic scale regime}
  resulting in a highly fluctuating fluid field.
\item[c)]  The interplay  of gravitational settling of dust grains and the
  convective up-mixing of uncondensed material  governs the {\it macroscopic scale regime}.
\end{itemize}

\begin{figure}[h]
\begin{center}
\epsfig{file=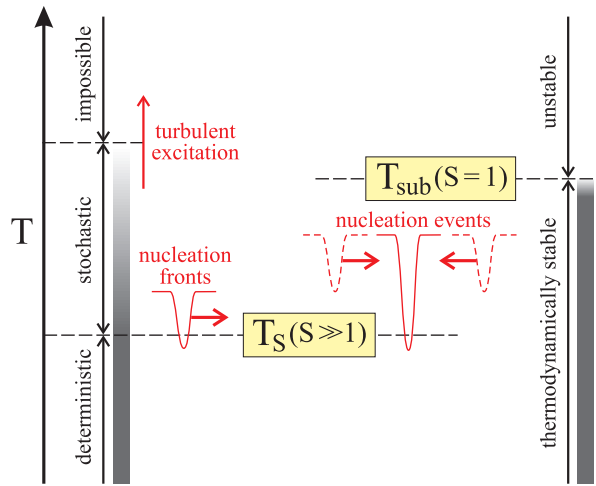, scale=0.6}
\caption[]{Regimes of turbulent dust formation. $T_S$: nucleation
  threshold temperature (supersaturation $S\ll 1$ required), 
  $T_{\rm sub}$: sublimation temperature.}
\label{fig:regimes}
\end{center}
\end{figure}

\noindent
The main scenario envisioned is a convectively ascending fluid element
in a brown dwarf atmosphere, which is excited by turbulent motions and
just reaches sufficiently low temperatures for condensation. The dust
formation in a turbulent gas is found to be strongly influenced by the
existence of a nucleation threshold temperature $T_S$. The local
temperature $T$ must at least temporarily decrease below this
threshold in order to provide the necessary supersaturation for
nucleation, e.g. by eddy interaction.

Depending on the relation between the local mean temperature
$\overline{T}$ and $T_S$, three different regimes can be distinguished
(see l.h.s. of Fig.~\ref{fig:regimes}): (i) the {\it deterministic}
regime ($\overline{T}<T_S$) where dust forms anyway, (ii) the {\it
  stochastic} regime ($\overline{T}>T_S$) where $T<T_S$ can only be
achieved locally and temporarily by turbulent temperature
fluctuations, and (iii) a regime where dust formation is {\it
  impossible}. The size of the stochastic regime depends on the
available turbulent energy. This picture of {\it turbulent dust
  formation} is quite different from the usually applied {\it
  thermodynamical picture} (r.h.s. in Fig.~\ref{fig:regimes}) where
dust is simply assumed to be present whenever $T<T_{\rm sub}$, where
$T_{\rm sub}$ is the sublimation temperature of a considered dust
material.

After initiation, the dust condensation process is completed by a
phase of active particle growth until the condensible elements are
consumed, thereby preserving the dust particle number density for long
times.  However, radiative cooling (as follow-up effect) is found to
have an important influence on the subsequent dust formation, if 
the dust opacity reaches a certain critical value. This cooling
leads to a decrease of $\overline{T}(t)$ which may re-initiate the
nucleation. This results in a runaway process (unstable feedback loop)
until radiative and phase equilibrium is achieved. Depending on the
difference between the initial mean temperature $\overline{T}(t\!=\!0)$
and the radiative equilibrium temperature $T_{\rm RE}$, a considerable
local temperature decrease and density increase occurs. Since the turbulent
initiation of the dust formation process is time-dependent and
spatially inhomogeneous, considerable spatial variations of all
physical quantities (hydro-, thermodynamics, dust) occur during the
short time interval of active dust formation (typically a few seconds
after initiation), which actually {\it creates} new turbulence.
Thus, small turbulent perturbations have large effects in dust
forming systems. 
Our 2D simulations show that the dust appears in lane-like and curled
structures. Small scale dust structures merge and form larger
structures. Vortices appear to be present preferentially in regions
without or with only little dust. Non of these structures would occur
without turbulent excitation.

From our work on small-scale turbulence simulations of dust-hostile
regions in substellar atmospheres, we compile necessary criteria for a
subgrid model of a dust forming, turbulent system.

\begin{center}
\colorbox{light}{\parbox{0.98\textwidth}{
\smallskip
\centerline{\underline{\bf Criteria on a subgrid model of a turbulent, dust-forming system:}}
\begin{itemize}
\item[\bf a)] The subgrid model should describe the transition
  stochastic $\rightarrow$ deterministic
  regime in dust forming turbulent fluids.

\item[\bf b)]  The dust formation process (nucleation + growth) is restricted
  to a short time interval since the dust formation time scales are
  much smaller than the large-scale hydrodynamic time scales.\\
  This involves that:
  \begin{itemize}
  \item[$\bullet$] The nucleation does occur only locally and event-like
    in very narrow time slots.

  \item[$\bullet$] The growth process continues as long as condensible material is
    available.
    
  \item[$\bullet$] The condensation process freezes in and the
    inhomogeneous dust properties are preserved. 

  \item[$\bullet$] Almost constant characteristic dust properties
    result in the mean-long term behavior.

  \end{itemize}
  
\item[\bf c)] The feedback loop with its fast radiative cooling should govern the
  transition from an almost adiabatic to an isothermal behavior of
  the dust/gas mixture.

\end{itemize}
}}
\end{center}

\noindent
The largest, observable scale of a brown dwarf atmosphere has been
investigated by attacking the drift problem.  A consistent theoretical
description was derived for dust formation and destruction,
gravitational settling, and element depletion including the effect of
convective overshoot. It was therewith possible for the first time to
overcome the widely used concept of ad hoc dust presents in substellar
atmospheres. Test calculations have shown that the dust will appear
stratified in the atmosphere causing a corresponding depletion of the
gas phase.

\medskip\noindent {\bf Acknowledgements:} This work has been supported
by the \emph{DFG} (grants SE 420/19-1,2; Kl 611/7-1; Kl 611/9-1) as part
of the DFG-Schwerpunkt {\it Analyse und Numerik von
  Erhaltungsgleichungen}.

\bibliographystyle{alpha}
\bibliography{/perm/PUB/literatur/ref801}
%


\printindex
\end{document}